\documentclass[prl,twocolumn,floats,aps,superscriptaddress,showpacs]{revtex4-1}  

\usepackage{amsmath}
\usepackage{graphicx}
\usepackage{amsfonts}
\usepackage{amssymb}
\usepackage{color}
\usepackage{footnote}

\usepackage{bm}% bo

\newcommand{\be}{\begin{equation}}
\newcommand{\ee}{\end{equation}}

\newcommand{\bea}{\begin{eqnarray}}
\newcommand{\eea}{\end{eqnarray}}

\begin{document}
 
\title{Response to ``Comment on Static correlations functions and domain
  walls in glass-forming liquids: The case of a sandwich geometry''
  [J. Chem. Phys. {\bf 144}, 227101 (2016)]}

\author{Giacomo Gradenigo}
\affiliation{Universit\'e Grenoble Alpes, LIPHY, F-38000 Grenoble, France and CNRS, LIPHY, F-38000 Grenoble, France \\}

\author{Roberto Trozzo}
\affiliation{Aix-Marseille Universit\'e, CNRS, Centrale Marseille, M2P2 UMR 7340, 13451, Marseille, France\\}

\author{Andrea Cavagna}
\affiliation{Istituto Sistemi Complessi, Consiglio Nazionale delle Ricerche, UOS Sapienza, 00185 Rome, Italy}
\affiliation{Dipartimento di Fisica, Universit\`a Sapienza, 00185 Rome, Italy \\}

\author{Tomas S. Grigera}
\affiliation{Instituto de Investigaciones Fisicoqu\'imicas Te\'oricas y Aplicadas (INIFTA) and Departamento de F\'isica,
Facultad de Ciencias Exactas, Universidad Nacional de La Plata, c.c. 16, suc. 4, 1900 La Plata, Argentina
CONICET La Plata, Consejo Nacional de Investigaciones Cientificas y Tecnicas, Argentina.\\} 

\maketitle

{\bf  The point-to-set correlation function has proved to be a very valuable
tool to probe structural correlations in disordered systems, but more
than that, its detailed behavior has been used to try to draw
information on the mechanisms leading to glassy behavior in
supercooled liquids. For this reason it is of primary importance to
discern which of those details are peculiar to glassy systems, and
which are general features of confinement. Thus the concerns raised
in~\cite{krako16} definitely need to meet an answer.} 

The Comment~\cite{krako16} proposes an alternative analysis of the
numerical data presented in \cite{US}, according to which the
behaviour of the point-to-set correlation function can be interpreted
as a linear superposition of boundary effects, rather than the effect
of non-trivial thermodynamics. We believe this alternative explanation
is not compelling.  The problem is that the expression Eq.~1 of
\cite{krako16} for the correlation $h_{\textrm{dis}}({\bf x},{\bf
  y})$, where $h_{\textrm{dis}}({\bf x},{\bf y})$ is the linear
superposition of the influence of the two boundaries, is based on the
assumption that at least one of the two points ${\bf x}$ or ${\bf y}$,
is ``far enough'' from an amorphous boundary. According to the
Comment, Eq.~1 suggests the general validity of a superposition
principle for the data of~\cite{US}.  What is puzzling is that the
most relevant information on non-trivial thermodynamics contained in
\cite{US} is related to narrow sandwiches, which is precisely the case
where Eq.~1 does not apply.  Therefore, since Eq.~1 only
\emph{suggests} the validity of a superposition principle and in
practice does not apply to the most important situation, i.e.\ narrow
sandwiches, it seems to us that there is no theory of simple liquids
behind the superposition principle, but just the assumption of a
reasonable physical mechanism. We find therefore quite unconvincing
the statement according to which Eq.~1 is the ``simple result''
previously not available which allowed to ``make concrete'' the
superposition scenario brought forward in~\cite{krako16}.  Hence, the
Comment provides an explanation of the non-exponentiality of the
point-to-set correlation function in the context of the 1D Ising
model. Let us remark to this purpose that there are several critical
phenomena taking place in simple liquids, like the demixing transition
in a binary mixture or the liquid-vapour transition in a monodisperse
system, which have universal features well captured by the physics of
the Ising model. Moreover, even some general ideas about the Random
First-Order Transition theory, for instance how the point-to-set
correlation \emph{should} behave changing the size of the confining
cavity in a system with short-range interactions, can be put under
scrutiny looking at the corresponding behaviour in magnetic systems,
see for instance~\cite{CC07} but also Sec. VI of ~\cite{US}. At the
same time we need to warn the reader that such a specific problem, as
whether the behaviour of the point-to-set correlation function for a
3D glass-forming liquid is due to trivial finite-size effects or to
thermodyamic anomalies, cannot be solved in favour of a ``simple
liquid scenario'' just looking at the behaviour of a 1D Ising model:
the latter shares too few commonalities with the system of interest
from the viewpoint of such a specific question. For instance, any
representation of a fluid in terms of an Ising model, cannot
distinguish between a simple fluid and a fluid with a complex energy
landscape. The ``simple liquid scenario'' cannot be supported solely
by the results on the 1D Ising in particular due the lack of any other
favourable evidence: we remark that the only theoretical formula
related to simple liquids (Eq.~1 of the Comment), does not hold in the
regime of narrow slits.

% Altri risultati:
Apart from this inconsistency, we would like to point out that results
more recent than \cite{US} provide clear evidence that non-trivial
thermodynamic behaviour occurs in a confined supercooled liquid.  Not
only theoretical models show that confinement, irrespectively to the
kind of boundary conditions~\cite{FM07,CGB13}, is already sufficient
to originate thermodynamic anomalies, but numerical evidence
independent from the behaviour of the point-to-set correlation
function clearly support the idea that these thermodynamic anomalies
are effectively present in the behaviour of confined supercooled
liquids ~\cite{MCG15,BCY16}: support for the existence of a
point-to-set correlation length $\xi_{\textrm{PS}}$ has also been
found from the peak of the specific heat of a confined supercooled
liquid as a function of the cavity size~\cite{MCG15}, as well as from
the distribution of the overlap as a function of the cavity
size~\cite{BCY16}, which at $\xi_{\textrm{PS}}$ is bimodal.

% Wrap up
In conclusion, considering all the numerical and theoretical
information about the behaviour of supercooled liquids in confined
geometries gathered in recent years, it seems to us that the RFOT
theory scenario remains still the most compelling explanation of the
non-exponential decay of the point-to-set correlation function first
discovered in \cite{BBCGV08}.

\end{document}